\documentstyle[12pt,epsf,epsfig]{article} 
\normalbaselines
\begin{document}
\bibliographystyle{unsrt}

\vbox {\vspace{6mm}} 

\begin{center}
{\large \bf EXPONENTIAL MOMENTS OF CANONICAL \\[2mm]
PHASE: HOMODYNE MEASUREMENTS}\\[7mm]
T. Opatrn\'{y}$^{\ast}$, M. Dakna,  and D.--G. Welsch\\
{\it Theoretisch-Physikalisches Institut, 
Friedrich-Schiller-Universit\"{a}t Jena 
\\
D-07743 Jena, Germany }
\end{center}

\vspace{2mm}

\begin{abstract}
A method  for  direct sampling
of the exponential moments  of canonical phase 
 from the data recorded in balanced 
homodyne detection is presented. Analytical expressions for the sampling 
functions are shown which are valid for arbitrary states. 
A numerical simulation illustrates the applicability of the method
and compares it with the direct measurement of phase 
by means of double homodyning.
\end{abstract}
 
\section{Introduction}

In the study of
the problem of phase of a quantum harmonic oscillator,
such as a radiation field, 
one usually proceeds  in one of two different ways.
 In the first, the phase is defined from the requirement that phase and 
photon number should be complementary quantities. This first-principle 
definition leads to the {\em canonical phase} 
related to the one sided unitary phase operator \cite{London,Susskind1}. 
In the second way,
phase quantities are defined from the output observed in phase-sensitive 
measurements, such as eight-port homodyne detection
\cite{Noh}. It was found that 
in such a scheme the $Q$ function 
is measured \cite{Walker}. The {\em measured phase} 
distribution can then be obtained by radial integration of the  
 $Q$ function. Whereas in the classical limit the measured 
phase coincides with the canonical phase, in the quantum regime the 
two phases significantly differ from each other in general.
The most important difference is that the integrated $Q$ function yields a
broader and less structurized distribution than the canonical phase
\cite{Leonhardt1}.
Is there a way which would bring together the advantages of these two
approaches - i.e., the theoretical elegance and pronounced structure 
typical of the
canonical phase and the advantage of
experimental availability as in the case for the integrated $Q$
function?

Before addressing the problem in more detail, let
us mention the concept of {\em direct sampling\/} of a quantity from
experimental data. This concept has been studied extensively in connection
with  the density matrix reconstruction \cite{Ariano1}. 
Assume a balanced homodyne measurement of a field quadrature
$x(\vartheta)$, where $\vartheta$ is the phase of the local oscillator
(LO). Performing such measurements with different $\vartheta$ 
on large ensembles of identically
prepared states we obtain 
probability distributions $p(x,\vartheta)$.
A quantity ${\cal A}$ can be sampled from the homodyne data if it can be
expressed as a two fold integral of the measured distribution,
\begin{eqnarray}
\label{c6a}
{\cal A} =\int_{2\pi} d \vartheta \int_{-\infty}^{\infty} dx ,
\, K_{\cal A}(x,\vartheta)\,p(x,\vartheta) ,
\end{eqnarray}
where $K_{\cal A}(x,\vartheta)$ is an integration kernel.
The quantity ${\cal A}$ can represent, e.g.,  density matrix elements, 
mean values of  operators, etc.  

Let us turn to the question of whether the canonical phase distribution can
be directly sampled. In this case the quantity  ${\cal A}$ in Eq. (\ref{c6a})
is the phase probability, ${\cal A}$ = $p(\varphi)$.
In Ref.  \cite{Dakna1} it was suggested to obtain 
the exact phase 
distribution 
as the limit of a convergent sequence of appropriately 
parametrized (smeared) distributions each of which can directly be 
sampled from the homodyne data. The exact phase distribution 
can then be obtained asymptotically to any degree of accuracy, if the 
sequence parameter is chosen such that smearing is suitably weak. 
However, it turned out that whereas the method works well for 
 states with low photon numbers, 
 with higher excitation the smearing parameter must be
chosen very small and the corresponding kernels  become more and more
structurized.
This makes  sampling problematic for highly excited
states. 
We can see a discrepancy
between the quantum and classical regions: on one
hand phase can easily be measured in classical physics, 
on the other hand 
sampling of the phase distribution becomes tedious
for  states from the classical
region. Can a unified approach be found which would 
bridge the gap between these two regions and which would 
enable us  to measure the canonical and classical phase distributions
on the same footing? 

In this contribution we show the possibility of direct sampling of 
the exponential phase moments 
of the canonical phase. The
corresponding kernels are well behaved functions which, for large $|x|$,
approach their classical counterparts. 
Since the moments contain the same
information about the phase properties as the probability distribution
itself, the method can serve as a way for experimental determination of
the canonical phase.


\section{Integration kernels}
\label{S2}

The canonical phase distribution $p(\varphi)$ 
of a state $\hat \varrho$ is defined as $p(\varphi)$
= $(2\pi)^{-1}$ $\langle \varphi | \hat \varrho | \varphi \rangle$, where
the phase states $|\varphi \rangle$ are \cite{Susskind1} 
$|\varphi \rangle$ = $\sum_{n=0}^{\infty} e^{in\varphi} |n\rangle $, 
$|n\rangle$ being the Fock states. The exponential phase moments 
$\Psi _{k}$ of this
distribution are given by
$\Psi _{k}$ = $\int_{2\pi}$ $e^{ik\varphi}$ $p(\varphi)$  $d\varphi$ 
and they can be expressed as 
\begin{eqnarray}
\label{qi1}
\Psi_k = \sum_{n=0}^{\infty}
\varrho_{n\!+\!k,n} 
\end{eqnarray}
for $k$ positive and $\Psi_k$ = $\Psi_{-k}^{\ast}$
for $k$ negative. Our aim is to express $\Psi_k$ by means of the measured
quadrature distribution $p(x,\vartheta)$, i.e., in the form of Eq.
(\ref{c6a}) with ${\cal A}$ $\equiv$ $\Psi_k$. For this purpose 
we must find the
corresponding kernel $K_{k}(x,\vartheta)$.

To do so, let us  express the distribution  $p(x,\vartheta)$ by means of
the density matrix elements as
\begin{eqnarray}
\label{qi3}
p(x,\vartheta) = \sum_{n\!=\!0}^{\infty} \sum_{m\!=\!0}^{\infty}
\psi_{n}(x) \psi_{m}(x)
\varrho_{m,n} e^{i(n-m)\vartheta} .
\end{eqnarray}
Here $\psi_{n}(x)$ are the eigenfunctions of the harmonic oscillator
Hamiltonian, $\psi_{n}(x)$ $=$ $(2^{n}n!\sqrt{\pi})^{-1/2}$ exp$(-x^{2}/2)$
H$_{n}(x)$, H$_{n}(x)$ being the Hermite polynomials.
Substituting Eq. (\ref{qi3}) into (\ref{c6a}) and comparing with Eq.
(\ref{qi1}) we find that the kernels must be of the form
$K_{k}(x,\vartheta)$ = $e^{ik\vartheta}$ $K_{k}(x)$, where the
$x$-dependent part must satisfy the integral equation
\begin{eqnarray}
\label{qi5}
2\pi\int_{-\infty}^{\infty} dx \,
K_{k}(x) \psi_{n\!+\!k}(x) \psi_{n}(x) = 1
\end{eqnarray}
($n$ $\!=$ $\!0,1,2,\dots$). In \cite{Dakna2}  the solution of
this equation is discussed in detail. Let us  write here the result in the form
\begin{eqnarray}
\lefteqn{
K_{2m}(x)=
\frac{m!}{(2\pi)^{m+1}}
\int_{-\infty}^{+\infty} dt_1 
\, e^{-t_1^2} \, \cdots
}
\nonumber \\ && \hspace{1ex} \times \, \cdots 
\int_{-\infty}^{+\infty} dt_{2m} \,
e^{-2mt_{2m}^2}
\,\bigg\{
\frac{\Phi[m+1,{\textstyle \frac{1}{2}}, 
z_{2m}(1+z_{2m})^{-1}x^2 ]}
{z_{2m}^m(1+z_{2m})^{m+1}}-\frac{1}{z_{2m}^m}
\bigg\} ,
\label{kernev}
\end{eqnarray}
\begin{eqnarray}
\lefteqn{
K_{2m+1}(x)=
\frac{2x(m+1)!}{(2\pi)^{m+3/2}}
\int_{-\infty}^{+\infty} dt_1 \, e^{-t_1^2} \, \cdots
}
\nonumber \\ && \hspace{1ex} \times \, \cdots
\int_{-\infty}^{+\infty}
dt_{2m+1} \, e^{-(2m+1)t_{2m+1}^2} \,
\frac{\Phi[m+2,{\textstyle \frac{3}{2}}, 
z_{2m+1}(1+z_{2m+1})^{-1}x^2]}
{z_{2m+1}^m(1+z_{2m+1})^{m+2}} \, ,
\label{kernod}
\end{eqnarray}
with $z_{k}$ $\!=$ $\![\exp(-\sum_{j=1}^{k}t_{j}^{2})-1]/2$,   
$\Phi (a,b,y)$ being the confluent hypergeometric function. 
\begin{figure}[tbh]
\begin{minipage}[b]{0.45\linewidth}
\centering\epsfig{figure=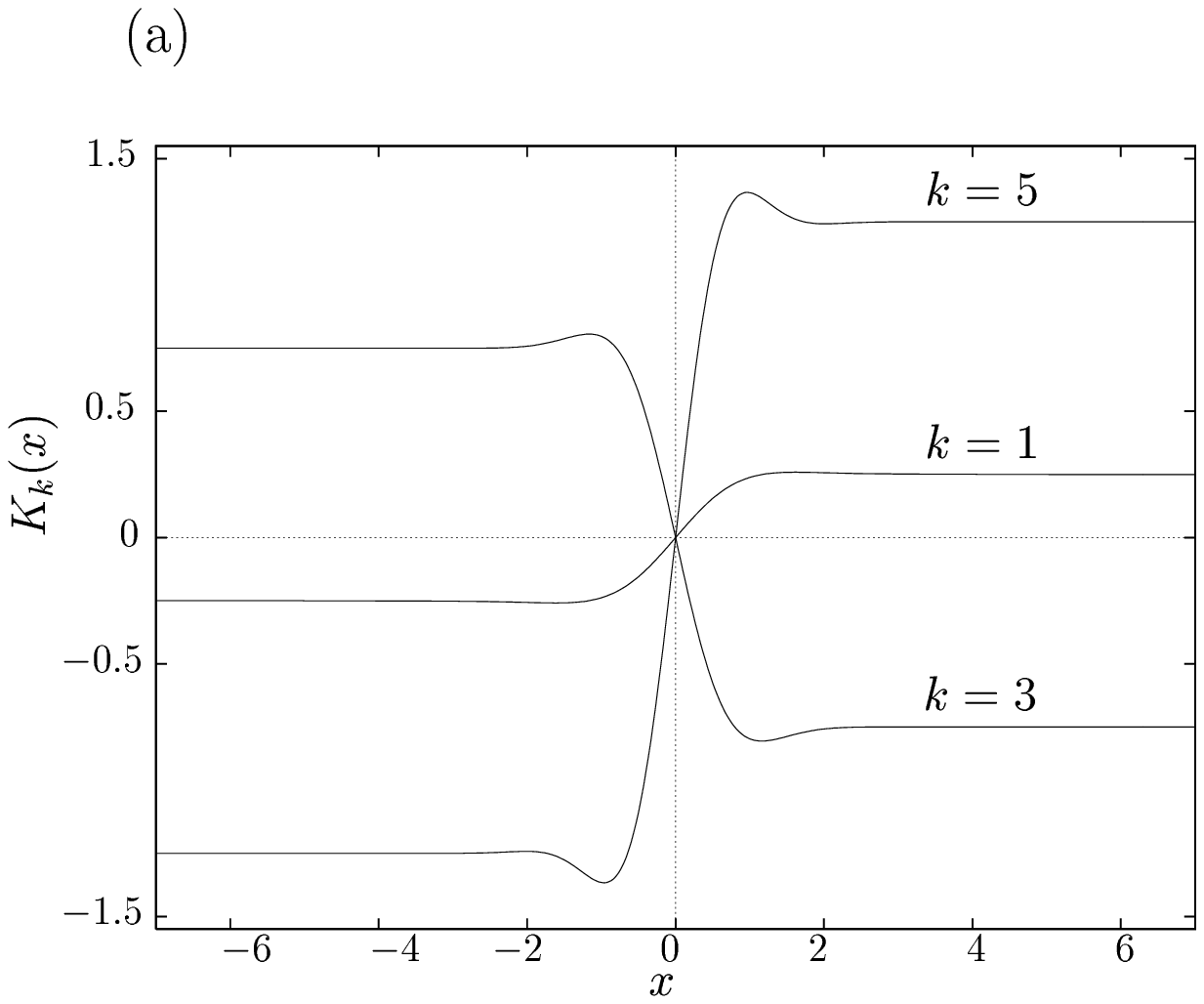,width=.9\linewidth}
\end{minipage} \hfill
\begin{minipage}[b]{0.45\linewidth}
\centering\epsfig{figure=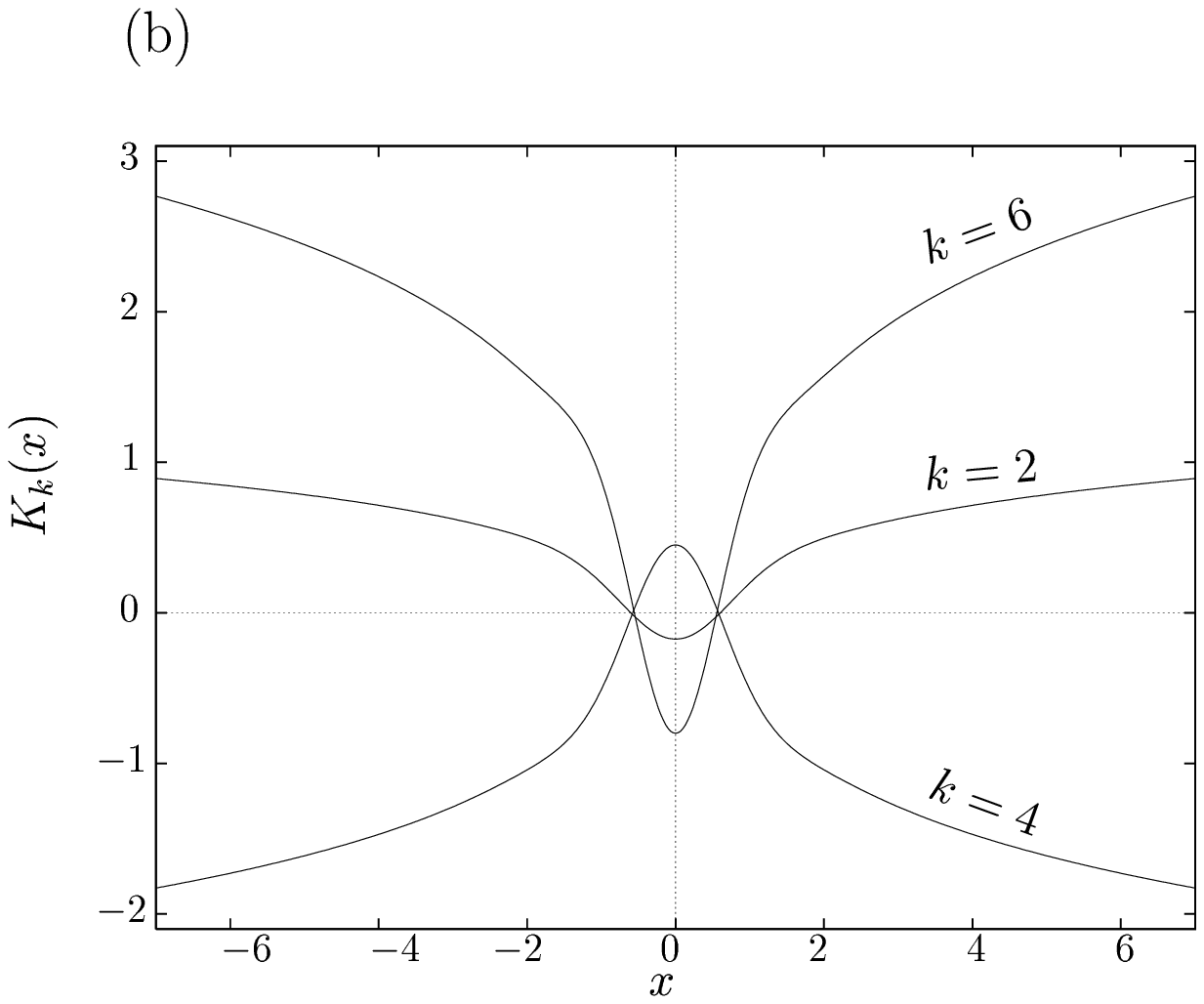,width=.9\linewidth}
\end{minipage} \hspace{1ex}
\caption{The functions $K_{k}(x)$.}
\end{figure}

In Fig. 1 we plot the functions $K_{k}(x)$ for several values of $k$. As
can be seen, the  functions $K_{k}(x)$ are well behaved and 
with increasing $|x|$ they quickly approach 
 their (classical) asymptotics $K_{k}^{c}(x)$, where
\begin{eqnarray}
\label{cl4}
K_{2m\!+\!1}^{c}(x) = {\textstyle\frac{1}{4}}
(-1)^{m} (2m+1)\, {\rm sign}\,(x)
\end{eqnarray}
and
\begin{eqnarray}
\label{cl5}
K_{2m}^{c}(x) = \pi^{-1} (-1)^{m+1} m \ln|x| + C_{2m},
\end{eqnarray}
$C_{2m}$ being a 
(unimportant) constant. 
An essential difference between $K_k(x)$ and $K_k^c(x)$ appears only for $x$
near zero, within the area of vacuum fluctuations.
It is shown in \cite{Dakna2} that the
functions  $K_{k}^{c}(x)$ can 
be obtained  as kernels for sampling
exponential phase moments in classical physics.
Therefore, we have found kernels for  direct sampling of the canonical
phase moments which can be used for any state, regardless if it has
typically quantum or classical properties.


\section{Simulated measurements}
\label{S3}
To illustrate the applicability of the  method we
have performed  computer simulations of homodyne measurements and used the
kernels $K_{k}(x,\vartheta)$ to determine the moments $\Psi_{k}$. 
For
comparison, we have also
simulated  double homodyne measurements to get exponential phase moments
that correspond to the radially integrated $Q$ function.
In the computer simulations the state to be detected is
the phase squeezed state $|\alpha,s \rangle$
with the coherent amplitude $\alpha$ = 5$\times e^{i\varphi_{0}}$,
$\varphi_{0}$ = 0.6, and the squeeze parameter $s$ = 6 (the mean
photon number of this state is $\langle n \rangle$ = 26.04). For both the
homodyne and the double homodyne measurements the total number of
measurement events is $N_{e}$ = 6020. For the homodyne measurements the
LO phase $\vartheta$ takes 41 values equidistantly 
distributed over the $2\pi$ interval. As shown in \cite{Dakna2},
 the statistical error of the  sampled 
phase moments depends on the numbers of measurement events for different
$\vartheta$. By a proper distribution of the total number  $N_{e}$ for
individual phases $\vartheta$ we can decrease the statistical error of
various moments $\Psi_k$. 
In order to minimize the statistical error of the first
moment $\Psi_{1}$, 
we increased the number of measurement events for such 
$\vartheta$ for which the peak of $p(x,\vartheta)$ is near  $x$ = 0
(maximum 800 events), whereas
for $\vartheta$ yielding a peak far from zero the number of events was small
(minimum 10 events). 
\begin{figure}[tbh]
\begin{minipage}[b]{0.45\linewidth}
\centering\epsfig{figure=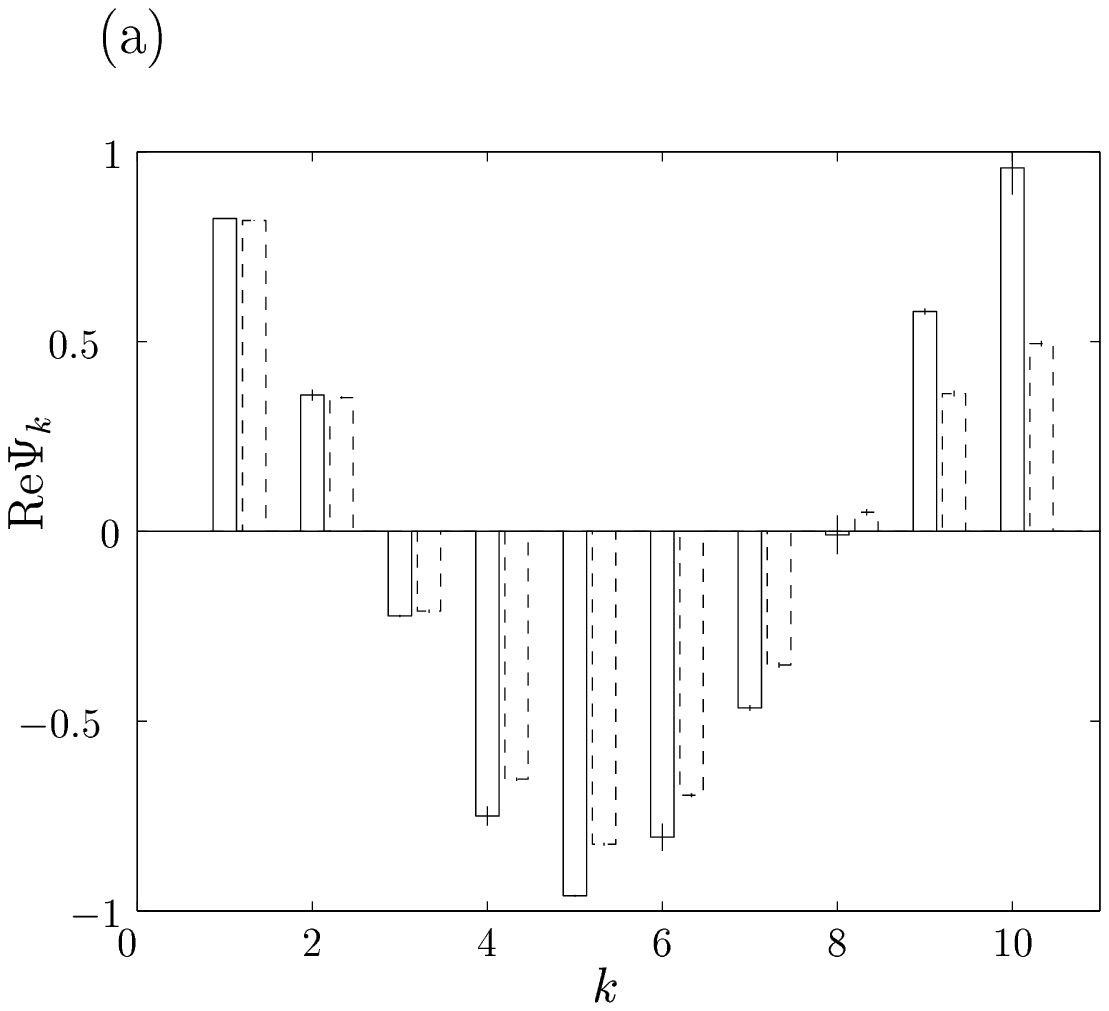,width=.9\linewidth}
\end{minipage} \hfill
\begin{minipage}[b]{0.45\linewidth}
\centering\epsfig{figure=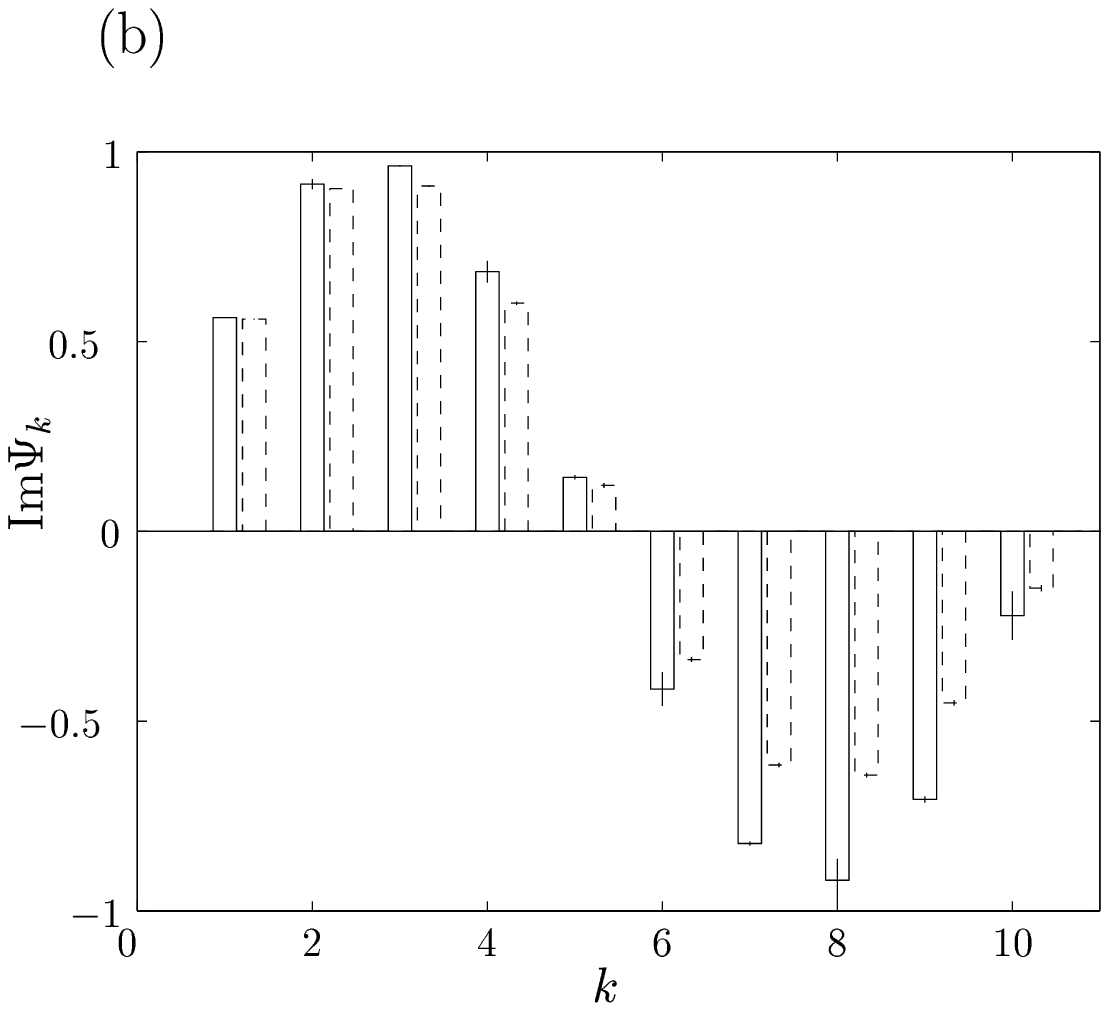,width=.9\linewidth}
\end{minipage} \hspace{1ex}
\caption{Real (a) and imaginary (b) parts of the 
experimentally determined exponential phase
moments $\Psi_{k}$. 
Bars with full lines: direct sampling from balanced homodyning, bars with 
dashed lines: double homodyning. The vertical lines represent the estimated
statistical error.}
\end{figure}

The experimentally determined phase moments together with 
the estimated statistical errors
are shown  in Fig. 2. We can see that the 
absolute values of moments of the integrated $Q$
function are smaller than the corresponding values of the canonical
distribution; the difference becomes larger with increasing $k$. This
corresponds to the fact that the integrated $Q$ function smears the 
structure of the canonical distribution. 

Let us compare the statistical errors. For the odd moments the errors are
approximately of the same magnitude
in the two methods, whereas for the even moments the sampling
method yields larger errors. This reflects the qualitatively
different behavior of the
kernels  for $k$ odd and $k$ even. 
It is also related to the chosen numbers of
measurement events for different $\vartheta$: in our example we have 
distributed the event numbers so as to minimize $\Psi_{1}$ 
which on the other hand increases the errors of even moments.
Even though one could expect that the double homodyning - as a direct phase
measurement - would yield generally smaller statistical errors, we find that
the statistical error of the first moment $\Psi_{1}$ is smaller for the
sampled canonical distribution. (The direct measurement means that a single measurement event
yields a single  value of phase.)

Let us mention that the first moment is connected 
to very important characteristics of the phase distribution.
The mean value of phase $\bar \varphi$ can be calculated
as $\bar \varphi$ = arg$\Psi_{1}$; 
this quantity can correspond, e.g., to a phase shift in an interferometer.
Since $\Psi_{1}$ is determined more precisely in the balanced homodyning than
in the double homodyning, the sampling method enables us to determine
$\bar \varphi$ with smaller statistical error.
 From the experimental data
 we obtain $\bar \varphi$ = 
0.5994$\pm$0.0011 for the sampling method, 
whereas from the integrated $Q$ function we obtain
$\bar \varphi$ =  0.5990$\pm$0.0016. (Note that for both distributions
the correct value is $\bar \varphi$ = $\varphi_{0}$ = 0.6.)
As can be seen, the error of determination of the
mean phase by means of homodyne sampling is
about 70\% of the error in the double homodyning.
The moment $\Psi_{1}$ is also related to
various phase uncertainties, which describe the
``width'' of the phase probability distribution. A phase uncertainty
$\Delta \varphi$ can be defined as $\Delta \varphi$ = arccos$|\Psi_{1}|$,
which is related to the Bandilla-Paul phase dispersion $\sigma^{2}_{BP}$ 
as $\sigma _{BP}$ = sin$\Delta \varphi$ and to the Holevo phase dispersion
 $\sigma^{2}_{H}$ 
as $\sigma _{H}$ = tan$\Delta \varphi$ \cite{Bandilla}. (An advantage of the
uncertainty $\Delta \varphi$ is that it enables us to measure the phase
width in the same units as the phase itself - in radians, degrees, etc.)
In this way we obtain $\Delta \varphi$ = 0.065 for the  canonical phase
distribution and $\Delta \varphi$ = 0.125 for the integrated $Q$ function.

\section{Discussion and conclusion}
\label{S6}

The presented method shows a very straightforward way for obtaining the
exponential moments of the canonical phase from the data of  homodyne
detection. Direct sampling enables us to reconstruct  the moments
$\Psi_{k}$ in real time as the experiment runs, together with the estimation
of the statistical error \cite{Dakna2}.
In this way the theoretically profound concept of canonical phase can be
connected with data obtained from present experiments.

The moments $\Psi_{k}$ contain the same information as the probability
distribution $p(\varphi)$. Therefore 
they can be used for reconstruction of the original function
$p(\varphi)$. However, even the lowest moments give us an interesting
information about the phase properties, e.g., 
the first moment $\Psi_{1}$ is
directly related to
 the mean value of phase and to the phase uncertainty.

The integration kernels are well-behaved functions which rapidly approach 
their asymptotics given either as step-functions (for odd moments) or 
logarithmic functions (for even moments). As shown in \cite{Dakna2}, these
functions can serve as kernels for sampling of the phase moments in
classical physics. 
Therefore, we have found a unified approach which connects the
measurement of the canonical phase with its classical counterpart.

It has been shown that the accuracy of the sampled moments of the
canonical phase is comparable with that of the directly measured radially
integrated $Q$ function. 
Moreover, when the total number of  measurement
events is the same,  the sampling method 
can yield the mean value of phase  more precisely
than the measurement of the $Q$ function. 
Also this aspect can make the presented
method very attractive for experimental applications.

\section*{Acknowledgments}
This work was supported by the Deutsche Forschungsgemeinschaft. 
We are grateful to G.M. D'Ariano, Z. Hradil, and V. Pe\v{r}inov\'{a} for
stimulating discussions.



\vspace{4ex}
\noindent
$^{\ast}\,$Permanent address:
Palack\'{y} University, Faculty of Natural Sciences,
Svobody~26, 77146 Olomouc, Czech Republic

\end{document}